\documentstyle[12pt]{article}
\input psfig
\newcommand{\be}{\begin{equation}}
\newcommand{\ee}{\end{equation}}
\newcommand{\ba}{\begin{eqnarray*}}
\newcommand{\ea}{\end{eqnarray*}}
\newcommand{\bfg}{\begin{figure}[hbtp]}
\newcommand{\efg}{\end{figure}}
\newcommand{\bver}{\begin{verbatim}}
\newcommand{\ever}{\end{verbatim}}
\newcommand{\bit}{\begin{itemize}}
\newcommand{\eit}{\end{itemize}}
\newcommand{\ben}{\begin{enumerate}}
\newcommand{\een}{\end{enumerate}}

\newcommand{\comment}[1]{}

\def\al{\alpha}

\begin{document}

\title{\bf{An Orientation Selective
Neural Network and its Application to Cosmic Muon Identification}}

\author{ \\   \\
{\bf Halina Abramowicz}\\
{\sl Deutches Elektronen-Synchrotron DESY, Hamburg}\\{\sl and}\\
{\sl School of Physics and  Astronomy, Tel Aviv University}\\
[0.4cm]
 {\bf David Horn, Ury Naftaly, Carmit Sahar--Pikielny }\\
{\sl School of Physics and  Astronomy, Tel Aviv University}\\
{\sl Tel Aviv 69978, Israel}
}

\date{ }
\maketitle

\vspace{0.5cm}

\begin{abstract}
  We propose a novel method for identification of a linear pattern of
  pixels on a two-dimensional grid.  Following principles employed by
  the visual cortex, we employ orientation selective neurons in
  a neural network which performs this task.  The method is
  then applied to a sample of data collected with the ZEUS detector at
  HERA in order to identify cosmic muons which leave a linear pattern
  of signals in the segmented uranium-scintillator calorimeter. A two
  dimensional representation of the relevant part of the detector is
  used. The results compared with a visual scan point to a very
  satisfactory cosmic muon identification. The algorithm performs well
  in the presence of noise and pixels with limited efficiency. Given
  its architecture, this system becomes a good candidate for fast
  pattern recognition in parallel processing devices.

\end{abstract}
\thispagestyle{empty}

\setcounter{page}{0}
\newpage
\section{Introduction}

A typical problem of experiments performed at high energy accelerators
aimed at studying novel effects in the field of Elementary Particle
Physics is the need to preselect interesting interactions at as early
a stage as possible, in order to keep the data volume manageable. One
class of events which have to be eliminated is due to cosmic muons
which pass all trigger conditions.

The most characteristic feature of cosmic muons is that they leave in
the detector a path of signals aligned along a straight line. 
 The efficiency of
pattern recognition algorithms depends strongly on the granularity
with which such a line is probed, on the level of noise and the
response efficiency of a given detector. Yet the efficiency of a
visual scan is fairly independent of those features.  This lead us to
look for a new approach through application of ideas from the field of
vision.

The visual cortex performs the difficult task of constructing the
image of the 3D external reality~\cite{Marr,Poggio} from the signals
which are based on 2D retinal images. Our problem is clearly
different, yet it has some similarities.  The information gathered in
a detector has finite and different resolution in different locations,
can be multivariate in nature (e.g. location and energy deposition),
and has empty regions where no signals come from.  Often the vision
problem is described as achieving a balanced integration of all
different inputs to obtain the general content of the picture. Here we
look for some characteristic detail in order to classify the event
into some particular category.  This calls for a differentiation of
the information, neglecting sometimes most of the available data and
retaining only some key features.

The main tool which we will borrow from the neuronal circuitry of the
visual cortex is the orientation selective simple cell~\cite{Hubel}. It
will be incorporated in the hidden layers of a feed forward neural
network, possessing a predefined receptive field with excitatory and
inhibitory connections.  Using these elements we develop a method for
identifying straight lines of varying slope and length on a grid with
limited resolution.  This method is then applied to the problem of
identifying cosmic muons in accelerator data, and compared with other
tools.

\section{Description of the Task}

This work was motivated by the observation that a visual scan, albeit
time consuming, is by far the most trustworthy and efficient way of
identifying
cosmic muon events, in this case in the analysis of electron proton
interactions with the ZEUS detector~\cite{ZEUSDET} at HERA. 

We try to employ a neural network (NN) which captures the key elements
of the visual scan.  We limit ourselves at this stage to cosmic muons
which enter the rear part of the ZEUS calorimeter in the direction
perpendicular to the beam-line. Because of a catastrophic energy loss
in the electromagnetic cells of the calorimeter they mimic an electron
trigger, which is the basis for selecting deep inelastic electron
proton interactions~\cite{ZEUSF2}.  It should be noted however that
these muons do not affect the physics analysis, their contamination is
estimated to be well below 1\%.

In a two-dimensional representation the granularity of the rear part
of the ZEUS calorimeter~\cite{ZEUSCAL} can be emulated roughly by a
$23 \times 23$ lattice of $20 \times 20$~cm$^2$ squares.  While such a
representation does not use the full information available in the
detector, it is sufficient for our study. In our language each cell of
this lattice will be denoted as a pixel. A pixel is activated if the
corresponding calorimeter cell is above a threshold level
predetermined by the properties of the detector.

\begin{figure}[hbtp]
\centerline{\psfig{figure=
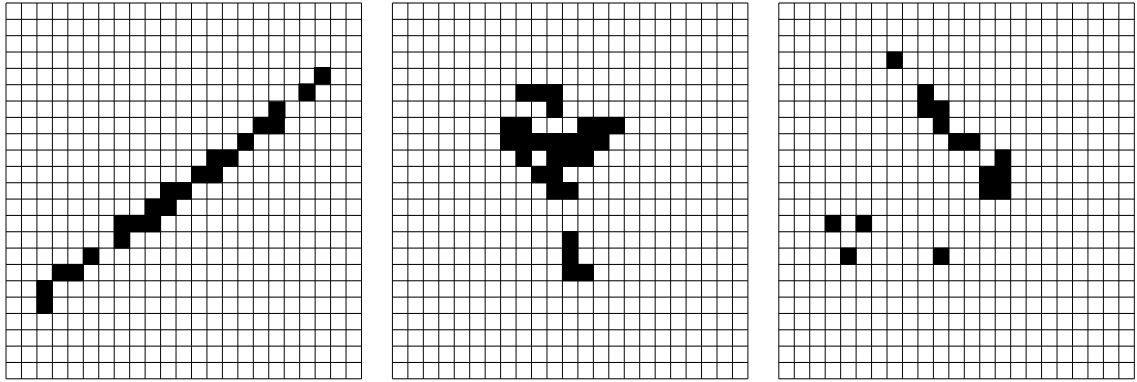,height=5.cm,width=15.cm}}
 \caption{Example of patterns corresponding to a cosmic muon (left),
 a typical accelerator event (middle), and an accelerator event which
looks like a muon (right), as seen in a two dimensional projection.}
\label{fig:muon_event}
\end{figure}

A cosmic muon, depending on its angle of incidence, activates along
its linear path typically from 3 to 25 neighboring pixels anywhere on
the $23 \times 23$ grid. The pattern of signals generated by
accelerator events consists on average of 3 to 8 clusters, of
typically 4 adjacent activated pixels, separated by empty pixels. The
clusters tend to populate the center of the $23 \times 23$ lattice.
Due to inherent dynamics of the interactions under study, the
distribution of clusters is not isotropic.  Examples of events,
as seen in the two-dimensional projection in the rear part of the ZEUS
calorimeter,
 are
shown in figure~\ref{fig:muon_event}.

The lattice discretizes the data and distorts it. Adding conventional
noise levels, the decision of classification of the data into
accelerator events and cosmic muon events is difficult to obtain
through automatic means.  Yet, it is the conventional feeling of
experimentalists dealing with these problems, that any expert can
distinguish between the two cases with high efficiency (identifying a
muon as such) and purity (not misidentifying an accelerator event).
We define our task as developing automatic means of doing the same.

\section{The Orientation Selective Neural Network }
Our analysis will be based on a network of orientation selective
neurons (OSNN) which will be described in this chapter.  We start
out with an input layer of pixels on a two dimensional grid with
discrete labeling $i=(x,y)$ of the neuron (pixel) which can get the
values $S_i=1$ or 0, depending on whether the pixel is activated or
not.

\begin{figure}[htbp]
\centerline{\psfig{figure=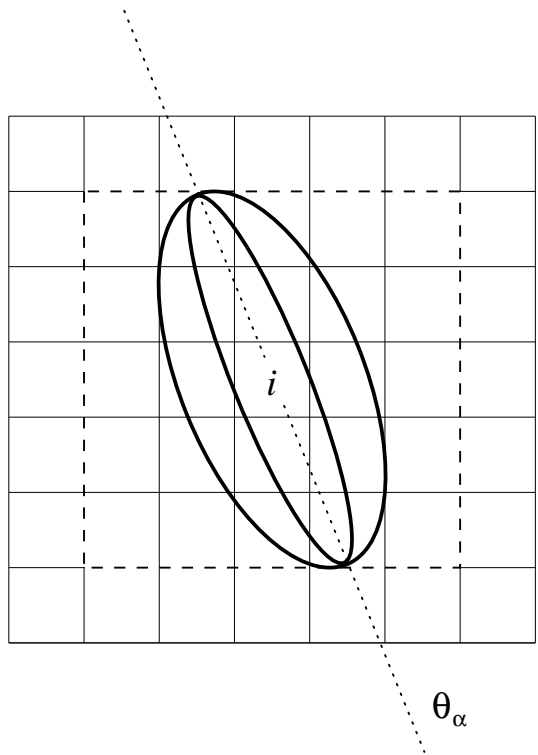,height=9.9cm,width=7.cm}}
\caption{The receptive field of a  $V^{i,\al}_2$ neuron centered
 around the pixel $i$
  with orientation $\theta_\al=101.5^{\circ}$.}
\label{fig:ellipse}
\end{figure}

The input is being fed into a second layer which is composed of
orientation selective neurons $V^{i,\al}_2$ at location $i$ with 
orientation $\theta_\al$ where $\al$ belongs to a discrete set of 16
labels, i.e. $\theta_\al=\al \pi/16$.  The neuron $V^{i,\al}_2$ is the
analog of a simple cell in the visual cortex.  The principle of its
design is presented in Figure~\ref{fig:ellipse}. Its receptive field
consists of an array of dimension $5\times 5$ centered at pixel $i$.
The angle $\theta_\al$ determines the orientation of a thin ellipse
centered at $i$ which encompasses an area of positive weights: each
neuron, $S_j$, at the input layer whose pixel is intersected by this
ellipse with a fractional geometrical overlap larger than $1/2$
excites $V^{i,\al}_2$ with weight $W_2^{i,\al,j}=1$.

The value of the weights $W_2^{i,\al,j}$ can be 1, 0 and -1. The
negative weights are attributed to the pixels which belong to the
$5\times 5$ subgrid but lie outside the fat ellipse, while the
positive values were assigned inside the thin ellipse.  All remaining
weights are 0.  Symbolically we may write the activation function of
the neuron on the second layer as,
\begin{equation}
V^{i,\al}_2={\cal F}_2( \sum_j W^{i,\al,j}_2S_j-T_2)\, ,
\label{eq1}
\end{equation}
where $T_2$ is a threshold value, and ${\cal F}_2(x) = (x+T_2)
\Theta(x)$. The values of $V^{i,\al}_2$ range from 0 to 5.  The second
layer consists then of $23\times23\times 16$ neurons, each of which
may be thought of as one of 16 orientation elements at some $(x,y)$
location of the input layer.  Next we employ a modified Winner Take
All (WTA) algorithm, selecting the leading orientation $\al_{max}(i)$
for which the largest $V^{i,\al}_2$ is obtained at the given location
$i$.  If we find that several $V^{i,\al}_2$ at the same location $i$
are close in value to the maximal one, we allow up to five different
$V^{i,\al}_2$ neurons to remain active at this stage of the
processing, provided they all lie within a sector of $\al_{max} \pm
2$, or $\theta_{max} \pm \pi/8$.  All other $V^{i,\al}_2$ are reset to
zero. If, however, at a given location $i$ we obtain several large
values of $V^{i,\al}_2$ which correspond to non-neighboring
orientations, all are being discarded.

The third layer also consists of orientation selective cells.
They are constructed with a receptive field of size $7\times7$,
and receive inputs from neurons with the
same orientation on the second layer. The weights $W_3^{i,\al,j}$
are defined in a similar fashion to $W_2^{i,\al,j}$.
The activation function of neurons of the
third layer is  
\begin{equation}
V^{i,\al}_3={\cal F}_3( \sum_j W^{i,\al,j}_3V^{j,\al}_2-T_3)\, ,
\label{eq2}
\end{equation}
where the index 3 stands for the quantities relevant to the third
layer. The sum over $j$ now runs over the larger receptive field.
$T_3$ is the threshold value for the third layer and ${\cal F}_3(x) =
(x+T_3) \Theta(x)$.  For linear patterns, the purpose of this layer is
to fill in the holes due to fluctuations in the pixel activation, i.e.
complete the lines of same orientation of the second layer.  As
before, we keep also here up to five highest values at each location,
following the same WTA procedure as on the second layer.

The fourth layer of the OSNN consists of only 16 components, $D^{\al}$, each
corresponding to one of the discrete orientations $\al$.  For each
orientation we calculate two global sums which are convolutions of
the first and third layers,
\begin{equation}
M^{\al}= \sum_i V^{i,\al}_3 S_i \qquad
N^{\al}= \sum_i V^{i,\al}_3 (1-S_i)\, . 
\label{eq3}
\end{equation}
The elements $M^{\al}$ carry the information about the number of the
input pixels that contribute to a given orientation $\theta_{\al}$
while $N^{\al}$ represent the mismatch between the neurons on the
third layer and the corresponding input pixels.

\begin{figure}[hbtp]
\centerline{\psfig{figure=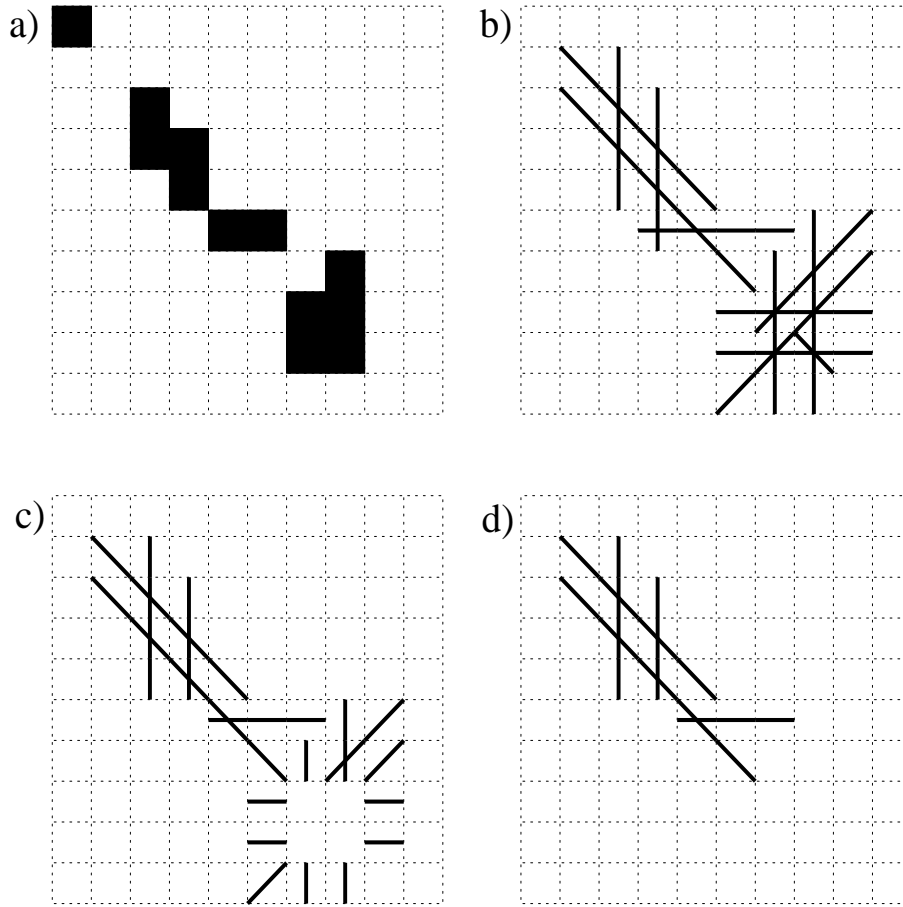,height=12.cm,width=12.cm}}
\caption{Demonstration of the OSNN filtering applied to the 
right most pattern presented in figure~\protect\ref{fig:muon_event}. 
a) A fragment of the lattice representing the input layer -- each small
square represents a pixel. b) Map of neurons which are activated in
the second layer. Each small square represents a neuron whose
orientation is designated by the line contained within the square. c)
Map of neurons surviving on the second layer after application of the
modified WTA algorithm. d) Neurons activated on the third layer.  }
\label{fig:event2}
\end{figure}

An example of the OSNN procedure applied to an accelerator event is
shown in Figure~\ref{fig:event2}.  The event is representative of a
class of patterns which are easily misidentified as cosmic muons.  For
the sake of simplicity we limit ourselves in this demonstration to
only four orientations, $\theta_{\alpha} = 0^{\circ},\,
45^{\circ},\,90^{\circ},\,135^{\circ}$, defined such that
$\theta_{\alpha} = 0^{\circ}$ corresponds to the horizontal and
$\theta_{\alpha} = 90^{\circ}$ to the vertical lines in the figure.
The convolution of the third layer with the input (fourth layer) leads
to the following results: \\
\begin{center}
\begin{tabular}{lcllcl}
$M_{0^{\circ}}$   & = & 2  &  $N_{0^{\circ}}$   & = & 1\\
$M_{45^{\circ}}$   & = & 0  &  $N_{45^{\circ}}$   & = & 0\\
$M_{90^{\circ}}$  & = & 4  &  $N_{90^{\circ}}$  & = & 3\\
$M_{135^{\circ}}$ & = & 5  &  $N_{135^{\circ}}$ & = & 4\\
\end{tabular}
\end{center}

The elements $D^{\al}$ defined as 
\begin{equation}
D^{\al}=M^{\al}-N^{\al}\, ,
\end{equation}
will serve as the basis for performing our task.
Cosmic muons are characterized by high values of $D^\al$ whereas
accelerator events possess low values. The simplest decision method
is to make a cut between the high and low values. A more sophisticated
approach is to use a simple neural network to perform the analysis
of the $D^\al$. Both methods will be discussed in the next chapter.

To explain the reason for the choice of the fourth layer, we have to
return to the origin of our analysis.  Since the orientation selective
elements act locally, and since accelerator events usually do not have
an overall orientation, we expect the orientation selective neurons on
the second and third layers to exhibit, for different angles, activity
patterns which do not resemble the input pattern.  On the other hand,
cosmic muons should generate, for the appropriate orientation,
activity patterns which resemble the input.

Because of the granularity and spatial resolution of our problem we
have to allow on the second layer the possibility of several
coexisting active neurons. This introduces a multiple working
hypothesis into the next steps to which the final decision is
deferred.  In the final calculational step we assign a figure of merit
which we want to maximize. This is built on the general intuition that
the strongest signal for a consistent straight line in one orientation
should be obtained from the orientation selective elements that lie on
this line in the data. Subtracting the effect of orientation neurons
which lie where no original pixels are excited, is needed in order to
discard accelerator events where accidentally $M^\al$ may be large and
yet the event is very different in shape from a single muon.

The complexity of this algorithm is ${\cal O}(n)$ where $n$ is the number of
pixels, since a constant number of operations is performed on each
pixel. There are basically four free parameters in the algorithm.
These are the size of the receptive fields on the second and third
layer and the corresponding activation thresholds. Their values can be
tuned for the best performance, however they are well constrained by
the spatial resolution, the noise level in the system and the
activation properties of the input pixels.  The size of the receptive
field determines to a large extent the number of orientations allowed
to survive in
the modified WTA algorithm.
\section{Training of the OSNN}

The details of the design of the OSNN and the tuning of its parameters
were fixed while training it on a sample of 250 cosmic muons and a similar
amount of accelerator events. The sample was obtained as a result of a
preselection with existing algorithms and a visual scan as a cross-check.

The size of the receptive fields was described in the previous
section. It is a compromise between a well defined orientation and the
length of a cosmic muon for which the algorithm becomes applicable.
The thresholds were chosen as $T_2=2.5 $ and $T_3=4.5 $.

\begin{figure}[hbtp]
\centerline{\psfig{figure=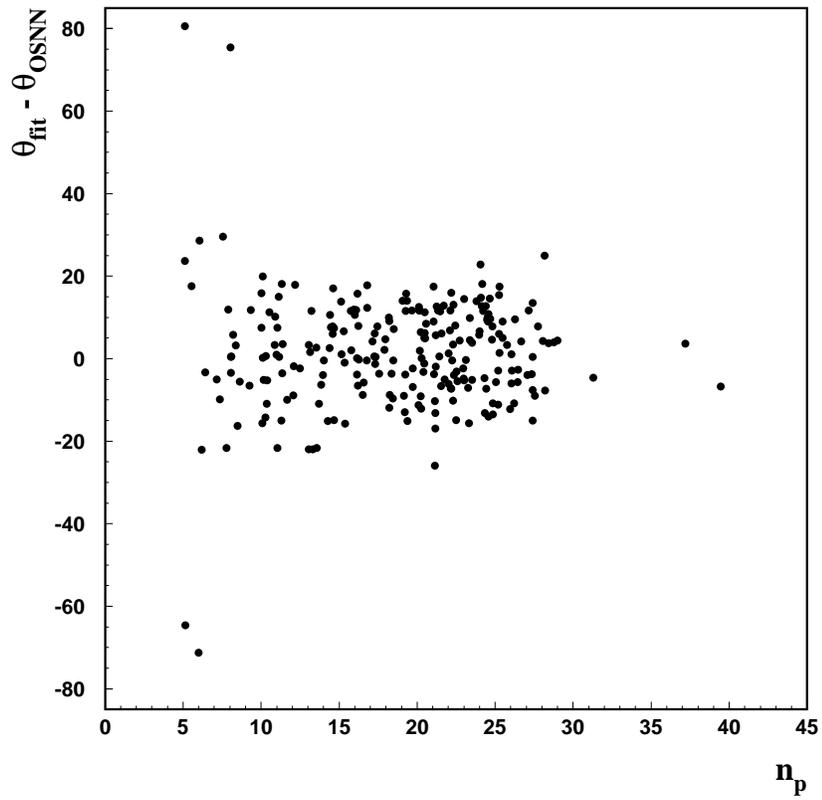,height=12.cm,width=12.cm}}
\caption{The difference between the
  muon direction as determined by the highest output from the OSNN,
  $\theta_{\rm OSNN}$, and the result of a straight line fit,
  $\theta_{\rm fit}$, as a function of the number of pixels activated
  in the input layer $n_p$.}
\label{fig:direction}
\end{figure}

For cosmic muon events the highest value of $D^\al$ determines the
orientation of the straight line, $\theta_{\rm OSNN}$. This
angle can be compared to the angle $\theta_{\rm fit}$, obtained from
a conventional straight line fit.  
The results are shown in figure~\ref{fig:direction} where we plot the
difference $\theta_{\rm fit}-\theta_{\rm OSNN}$ as a function of the
number $n_p$ of input pixels. We observe a good correlation between
the two angles. The spread in $\theta_{\rm fit}-\theta_{\rm OSNN}$
reflects the spatial resolution of the lattice and justifies the use
of the modified WTA with an angular uncertainty of $\pm \pi/8$. The
small contribution of erroneous results observed for very low values
of $n_p$ is not surprising in view of the sizes of our ellipses.

\begin{figure}[hbtp]
\centerline{\psfig{figure=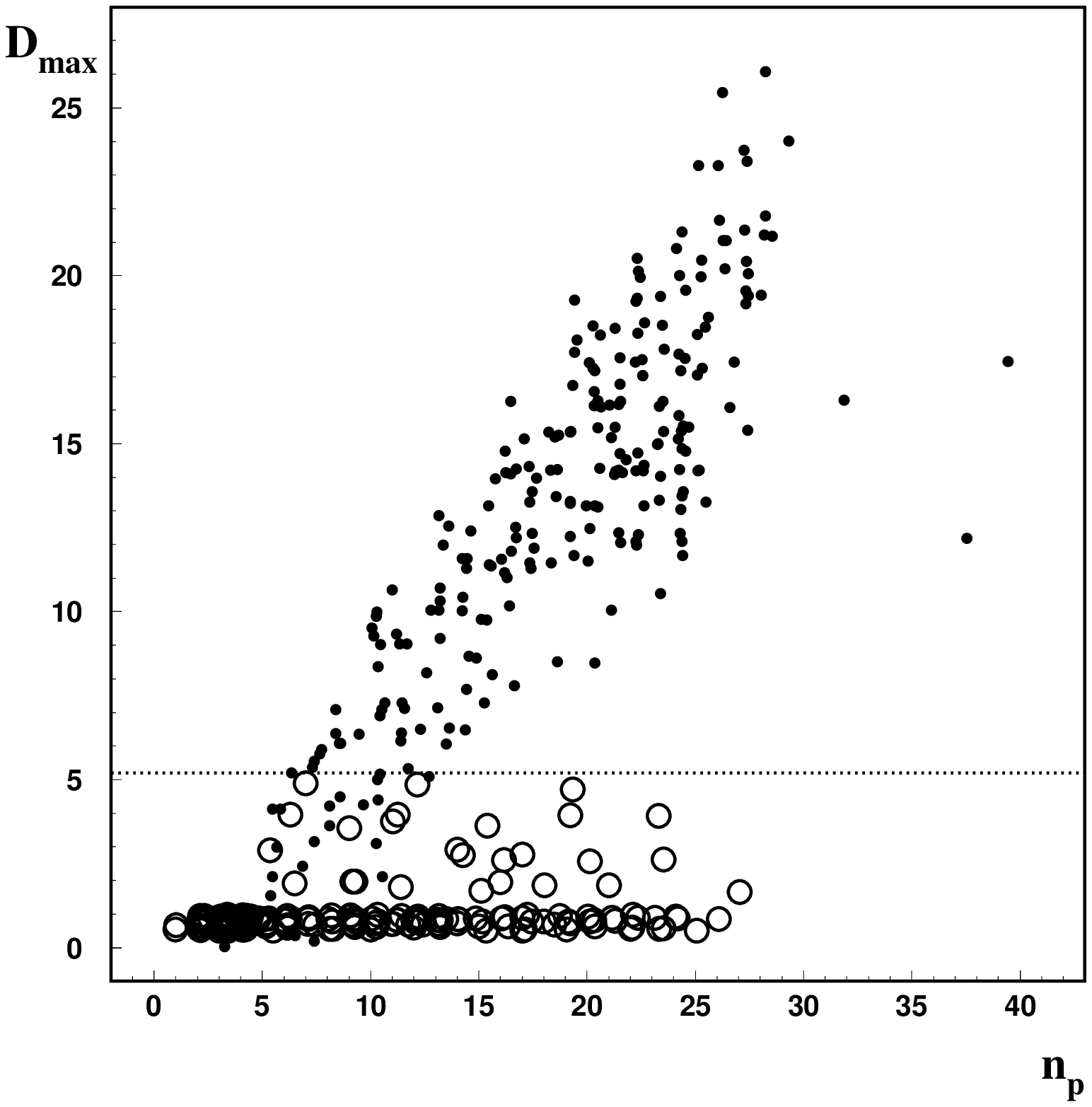,height=12.cm,width=12.cm}}
\caption{Correlation between the maximum value of $D^\al$, 
  $D_{\rm max}$, and the number $n_p$ of input pixels for cosmic muon
  (dots) and accelerator events (open circles). The dashed line defines
  a separator such that all events above it correspond to cosmic muons
  (100\% purity). This selection criterion has 92.2\% efficiency.}
\label{fig:gen}
\end{figure}

In figure~\ref{fig:gen} we present the correlation between the maximum
value of $D^\al$, $D_{\rm max}$, and the number $n_p$ of input
pixels for cosmic muon and accelerator events. As expected one
observes a linear correlation between $D_{\rm max}$ and $n_p$ for
the muons while almost no correlation is observed for accelerator
events.
Above $D_{\rm max}=5$ there is a very clear separation between
cosmic muon and accelerator events. This will define our first  separation
procedure which will be called OSNN-$D$. 

We will quantify the quality of the selection by quoting the
efficiency of properly identifying a cosmic muon for 100\% purity,
corresponding to no accelerator event misidentified as a muon. 
In the OSNN-$D$, where we require 
for a cosmic muon $D_{\rm max} \geq 5$, we achieve 92.2\% 
efficiency.

A better separation can be
achieved with more sophisticated selection criteria which use more
information from the fourth layer. Alternative methods
will be discussed below. Clearly such procedures 
will be affected  by the type of backgrounds to the linear
patterns that one wants to isolate.

\section{The Hough Transform}

A well known approach to detection of lines in an image is the
application of the Hough transform~\cite{Hough,Ballard,Duda}.
This method consists of a preliminary conversion of the Cartesian space
into a new parameter space. A line can be parameterized by two values
$\theta$ and $r$ according to the following expression:
\begin{equation}
r=x\cos\theta-y\sin\theta.
\label{rtheta}
\end{equation}
 A point $(x_i,y_i)$ in the Cartesian plane transforms into a curve in
the $(r,\theta)$ plane corresponding to all possible lines passing
through this point. An intersection point $(r_k,\theta_k)$ of two
curves in the $(r,\theta)$ plane indicates that the two points lie on
a straight line in the Cartesian plane. Thus, several points lying on
a straight line in the Cartesian plane produce an high-order intersection
point in the $(r,\theta)$ plane. 

When the straight line is probed with a finite resolution, like in the
case we are studying, one has to resort to a coarse graining of the
parameter plane and to search for       
an {\it intersection region} rather than an
intersection point. 
 To each pixel
activated in the input layer of our Cartesian lattice we assign
coordinates corresponding to the geometrical center of the pixel
$(x_i, y_i)$ that determine a curve as given
by equation~\ref{rtheta}. We discretize the $(r,\theta)$ space into bins
and 
count the number of curves that pass through them. The size of the bins
is optimized to give the best sensitivity to linear patterns.

\begin{figure}[hbtp]
\centerline{\psfig{figure=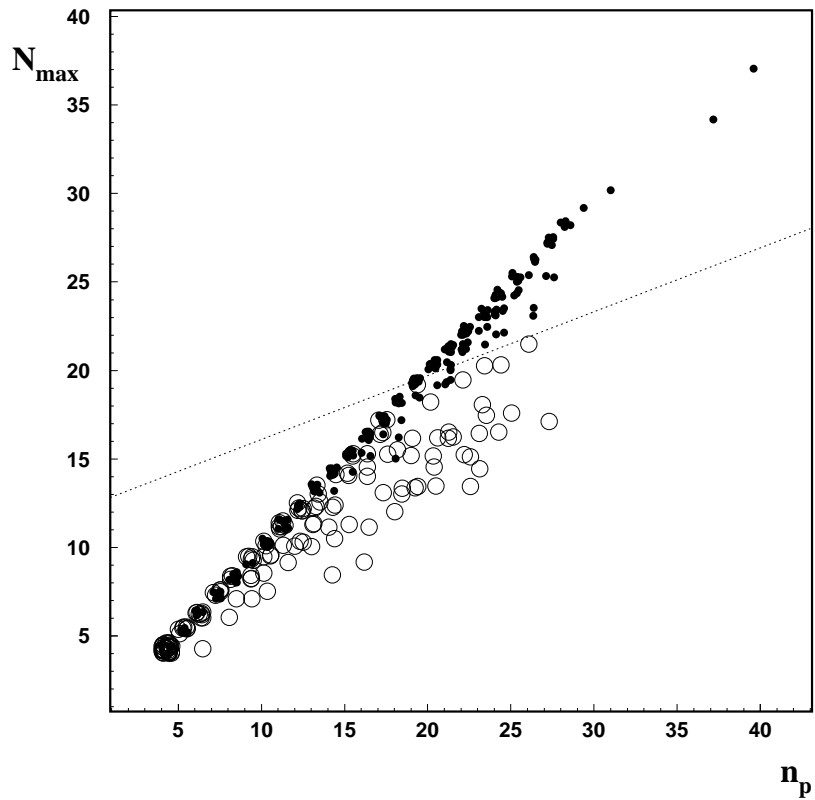,height=12.cm,width=12.cm}}
\caption{Correlation between the Hough value of 
  $N_{\rm max}$, and the number $n_p$ of input pixels for cosmic muon
  (dots) and accelerator events (open circles) The dashed line defines
  a separator such that all events above it correspond to cosmic muons
  (100\% purity). This selection criterion has only 47\% efficiency.}
\label{fig:hough_results}
\end{figure}
For each event, cosmic muon and accelerator, we determine the maximum
number of lines that cross one bin -- $N_{\rm max}$. In
figure~\ref{fig:hough_results} we present the distribution of
$N_{\rm max}$ as a function of $ n_p$ for the two classes of events.
 As expected, $N_{\rm max}$
is on the average larger for cosmic muons than for accelerator events.
However we observe that the overlap between the two classes of events
is very large. In fact for a 100\% purity of the cosmic-muon sample
the efficiency is as low as 47\%. This can be achieved by the linear
separator indicated on the figure. Obviously this result is much
worse than the OSNN-$D$ method described in the previous Section.
Note that in figure~\ref{fig:gen} the accelerator events were characterized
by low $D$ values which did not rise as function of $n_p$, thus allowing
for a good separation. Regrettably, this is not the case for
 $N_{\rm max}\,\, vs\,\, n_p$.

There are two reasons why the Hough transform performs poorly in this
context. The first one is that our linear patterns are thick. In
order to observe a definite enhancement in the $(r,\theta)$ space we have
to allow for many $(x_i,y_i)$ lying on a thin line to contribute. We thus
loose sensitivity to thick linear patterns
which include some natural jitter on the square lattice.
The other
reason is that the Hough transform is not directly sensitive to
the inactive pixels lying on a straight line. This is to say that
three non consecutive pixels lying on a straight line will look the
same as three consecutive ones in the $(r,\theta)$ space. Both 
these features are properly handled by the OSNN.

\section{OSNN Selection Procedures}

In an effort to improve further the performance of the OSNN we have
looked into the possibility of using more information from the fourth
layer. In particular, taking into account the angular resolution of
the orientation filters we selected the three leading $D^\al$ and
looked at the sum of the three terms.  For a 100\% purity the
efficiency improved slightly and became 93.7\%.

If instead of applying a simple cut we use a neural network to search
for the best classification of events with the OSNN outputs, we obtain
still better results.  The auxiliary network has 6 inputs, one hidden
layer with 5 nodes and one output unit.  The input consists
of a set of five consecutive $D^\al$ centered around $D_{\rm max}$
and the total number of activated input pixels, $n_p$. The cosmic
muons are assigned an output value $s=1$ and the accelerator events
$s=0$. The net is trained on our sample with error back-propagation.
This results in an improved separation of cosmic muon events from the
rest. In particular for $s \geq 0.6$ no accelerator events are found
and the muons are selected with an
efficiency of 96\%. This selection procedure
will be denoted as OSNN-$S$.

\begin{figure}[hbtp]
\centerline{\psfig{figure=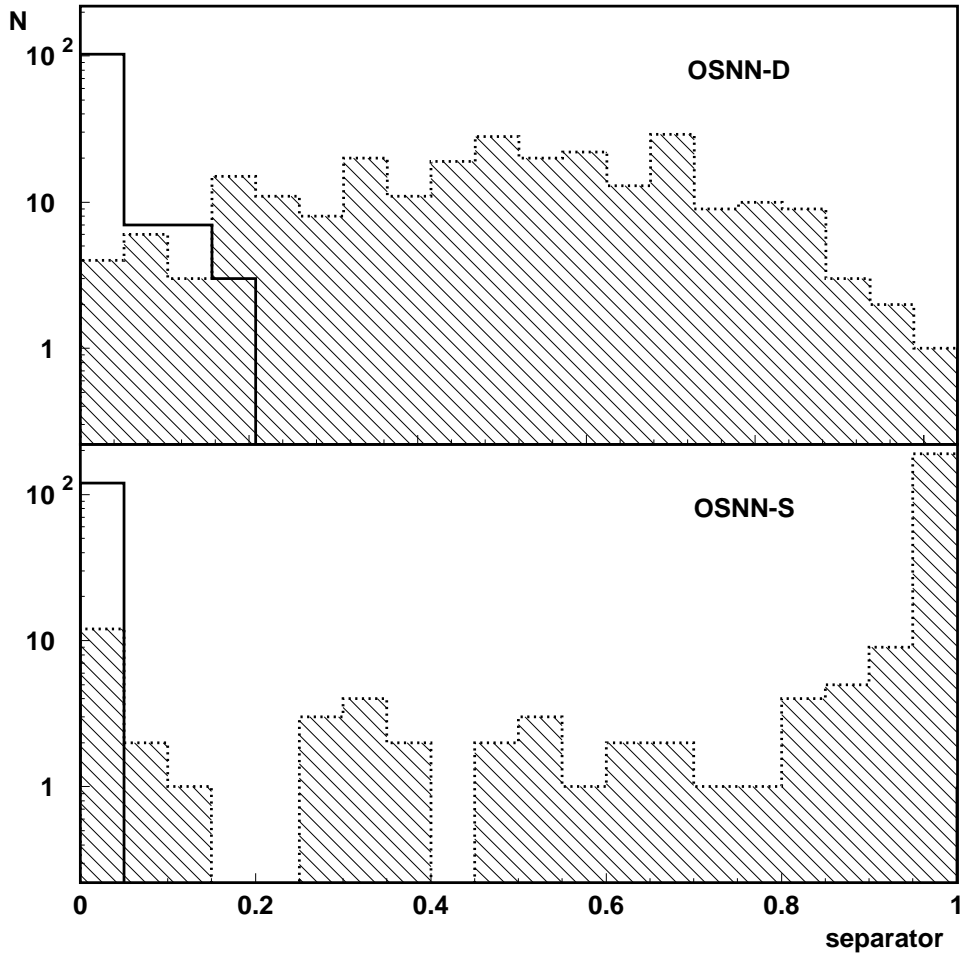,height=15.cm,width=15.cm}}
\caption{
Comparison of two OSNN separation methods: OSNN-$D$, based on the
$D_{\rm max}$ criterion, and OSNN-$S$, based on the auxiliary neural
network. For the sake of comparison the separator for OSNN-$D$ has
been defined as $D_{\rm max}$ rescaled by the maximum value achieved
for the training sample. For OSNN-$S$ the separator is defined as the
output of the auxiliary neural net $s$.  Accelerator events are
depicted by the solid line, while muons are represented by the shaded area.}
\label{fig:separ3}
\end{figure}
In order to compare the separation capability of OSNN-$D$ and
OSNN-$S$ we present in figure~\ref{fig:separ3} the distribution of the
cosmic-muon and accelerator events of the training sample in each of
the respective variables, $D_{\rm max}$ and $s$.
For the presentation  $D_{\rm max}$ has been rescaled
by the maximum value it achieved on the training
sample. The OSNN-$S$ is slightly better than  OSNN-$D$ in that there is
a better separation of the two classes of events.

\section{Performance of OSNN-$S$ on a Test Sample}

\begin{figure}[hbtp]
\centerline{\psfig{figure=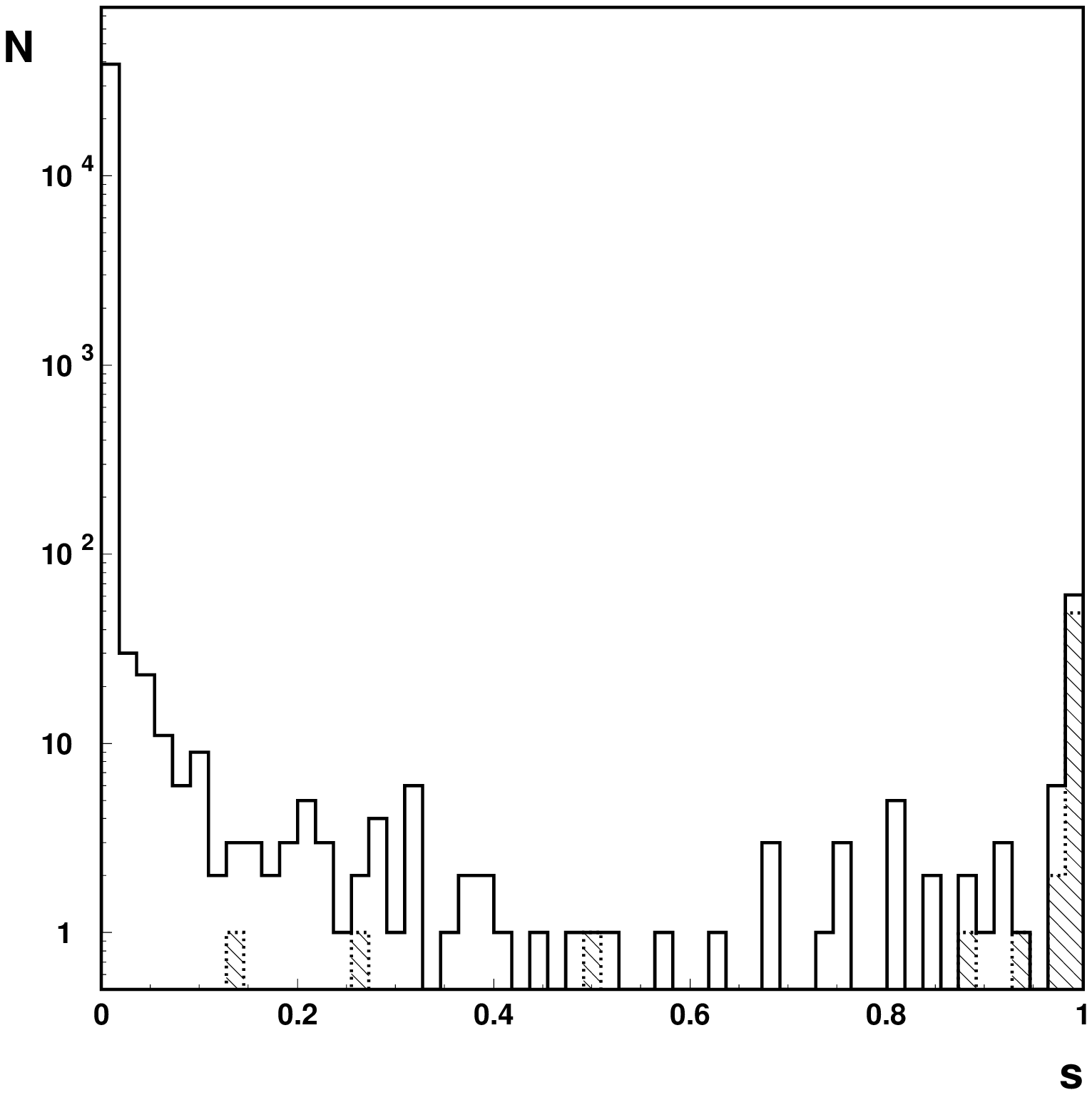,height=12.cm,width=12.cm}}
\caption{Distribution of the auxiliary neural network output $s$
  obtained with the OSNN-$S$ selector for the test sample.
  Cosmic muons are represented by the shaded areas.}
\label{fig:test1}
\end{figure}

Defining our final procedure as the OSNN with the auxiliary neural
network, OSNN-$S$, we apply it to a sample of 39,244 data events which
passed the standard physics cuts~\cite{ZEUSF2}. The distribution of
the neural network output $s$ is presented in figure~\ref{fig:test1}.
It looks very different from the one obtained with the training sample.
Whereas the former consisted of approximately 500 events distributed
equally among accelerator events and cosmic muons, this one contains
mostly accelerator events, with less than 1\% of muons.
This proportion is characteristic of  physics samples.
The distribution is approximately exponential with
a long tail towards larger values of $s$ and a small enhancement at
$s=1$.

We first perform a visual scan of all 138 events with $s \geq 0.1$.
The cut is motivated by the performance on the training sample. The
scan is based on the full information from the detector. The results
are shown in figure~\ref{fig:test1}, where the cosmic-muon events
are represented by the shaded area. We find 56 cosmic-muon events and
82 accelerator events. As expected the muons populate mainly the
region of large $s$ values.

Having now the benefit of a huge sample we see that
 the tail of the accelerator events spreads
to much larger values of $s$ than in the training sample.
It should be emphasized that this test sample is judged by different
criteria from those used on the training sample. Here cosmic muons or 
accelerator events are identified as such by the information which is
available from the whole detector, which is much more than the two
dimensional projection of its rear part which is used as an input
to the OSNN.
 A visual scan based solely on the two-dimensional
representation of the rear part of the ZEUS calorimeter reveals that
51 out of the 82 accelerator events look like short cosmic-muon
events.

For the sake of completeness we have also looked into the sample of
events with $s<0.1$. Using the full information from the detector
supplemented by a visual scan of events likely to be cosmic muons
we
find 15 cosmic-muon events. Most of these events would be classified
as accelerator events based on the information that is available to
the OSNN.

\begin{figure}[hbtp]
\centerline{\psfig{figure=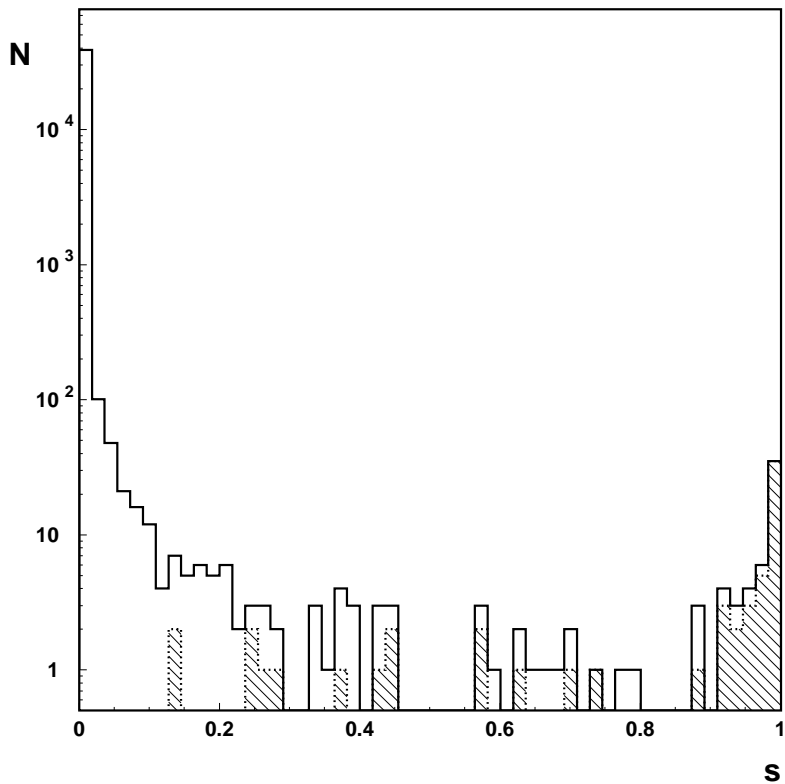,height=12.cm,width=12.cm}}
\caption{Distribution of the auxiliary neural network output $s$
  obtained after retraining the network on the test
sample. Cosmic muons are represented by the shaded areas.}
\label{fig:test2}
\end{figure}

Keeping our basic OSNN structure, we decided to use the new sample of
138 events with $s\geq 0.1$ for retraining the auxiliary network. 
This was then applied to the whole sample of 39,244 events, resulting in
a more pronounced bimodal distribution, as shown in
figure~\ref{fig:test2}. This suggests using 
$s=0.5$ as a separation point. We find then
54 out of the 71 muons of the total sample. This domain is
still contaminated by 15 accelerator events, 13 of which resemble
muons on the input layer of the OSNN. For comparison, before retraining 37
accelerator events were found for $s\geq 0.5$.
The retraining procedure shifted a few of the high-$s$ accelerator
events to lower $s$ values, thus leading to a better
separation between cosmic-muon and accelerator events.

\begin{figure}[htbp]
\centerline{\psfig{figure=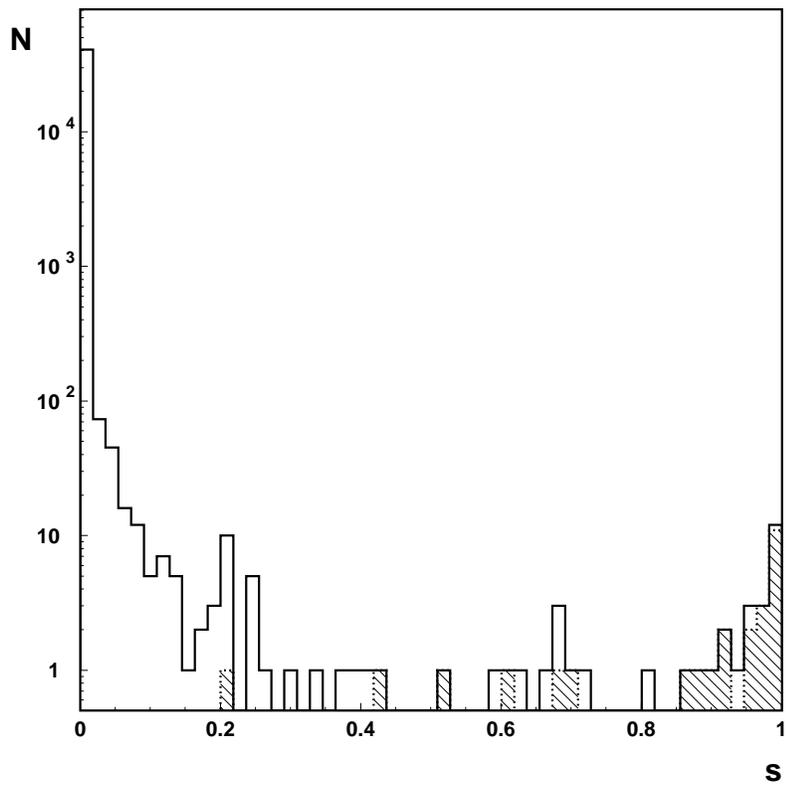,height=12.cm,width=12.cm}}
\caption{Distribution of the auxiliary neural network output $s$
  obtained with the retrained OSNN-$S$ selector for the new test
sample.  Cosmic muons are represented by the shaded areas.}
\label{fig:test3}
\end{figure}

To check the
stability of our procedure we apply the retrained OSNN-$S$ to a new
test set.  The result is displayed in figure~\ref{fig:test3}.  This
set contains 40,886 events, and its $s$ distribution is similar to the
retraining set of figure~\ref{fig:test2}.  We repeat the scanning
procedure described above and obtain very similar results. The cosmic
muons found with the full detector information for events with $s\geq
0.1$ are displayed by the shaded area. For $s\geq 0.1$ we find 27
cosmic-muon events and 51 accelerator events, out of which 16 look
like cosmic-muons at the input of the OSNN.
For $s\geq 0.5$ we find 10 accelerator
 events contaminating a sample of 25 muons,
however 8 out of the 10 events look like cosmic-muons at the input of
the OSNN.
Interestingly enough, a visual scan of 122 events below
$s=0.1$, selected as possible muon candidates because only the rear
part of the detector was activated, revealed no further muons.

We learn therefore that our method is stable and reproducible.  Its
remaining imperfection is due to the fact that the information that is
fed into the OSNN comes from only one part of the detector.  Even
with all its limitations it reduces the problem of rejecting
cosmic-muon events down to scanning a fraction of a percent of all the
events. We conclude that we have achieved the goal that we set for
ourselves, that of replacing a laborious
 visual scan by a computer algorithm
with similar reliability.

\section{Summary}

We have presented an algorithm for identifying linear patterns on a
two-dimensional lattice based on the concept of an orientation
selective cell, a concept borrowed from neurobiology of vision.
Constructing a multi-layered neural network with fixed architecture
which implements orientation selectivity, we define output elements
corresponding to different orientations, which allow us to make a
selection decision. The algorithm takes into account the granularity
of the lattice as well as the presence of noise and inefficiencies.

It has been applied successfully to a sample of events collected with
the ZEUS detector at HERA. We find a high efficiency and purity for
identifying cosmic muon events which leave a linear pattern of signals
in the rear part of the ZEUS calorimeter.

Since we use a fixed architecture, the complexity of our OSNN is not
very high. It has, though, a relatively large number of elements,
which increases proportionally to the number of pixels and the number
of orientations which is appropriate for a given analysis.  Such an
architecture is suitable for hardware implementation, in which case it
can provide for very fast parallel computation.

\subsubsection*{Acknowledgements}
We are indebted to the ZEUS Collaboration for allowing us to use the
sample of data for this analysis.  This work was partly supported by
the Israel Science Foundation.

 \end{document}